\documentclass[showpacs,prl,twocolumn]{revtex4}
\usepackage{graphicx}
\begin{document}
\title{Dynamic self-assembly and patterns in electrostatically
driven granular media}
\author{M. V. Sapozhnikov$^{1,2}$, Y. V. Tolmachev$^1$, I. S. Aranson$^1$ and W.-K. Kwok$^1$}
\affiliation{$^{1}$Materials Science Division, Argonne National
Laboratory, 9700 South Cass Avenue, Argonne , IL 60439 \\
$^{2}$Institute for Physics of Microstructures, Russian Academy of Sciences, GSP-105,
Nizhny Novgorod, 603000, Russia }
\date{\today}
\begin{abstract}
We show that granular media consisting of metallic microparticles
immersed in a poorly conducting liquid  in strong DC electric
field self-assemble a rich variety of novel phases. These phases
include static precipitate: honeycombs and Wigner crystals; and
novel dynamic condensate: toroidal vortices and pulsating rings.
The observed structures are explained by the interplay between
charged granular gas  and electrohydrodynamic  convective flows in
the liquid.
\end{abstract}
 \maketitle

Understanding the unifying principles of self-assembly of complex
systems such as macro-molecules \cite{winfree}, diblock copolymers
\cite{lopes}, micro-magnetic systems \cite{woods}, ensembles of
charged particles \cite{hayward} is the key to future advances in
nanoscience. Large ensembles of small particles display
fascinating collective behavior when they acquire an electric
charge and respond to competing long-range electromagnetic and
short-range contact forces. Many industrial technologies face the
challenge of assembling and separating such single- or
multi-component micro and nano- size ensembles.  The dynamics of
conducting microparticles in  electric field in the air was
studied in \cite{ar1,ar2}. Phase transitions and clustering
instability of the electrostatically driven granular gas were
found. The studies of self- assembly of colloidal particles in
aqueous solutions revealed the importance of self-induced
electrohydrodynamic (EHD) convective flows on the formation of
various precipitate states \cite{hayward,trau,yeh,solom,nadal}.
Ordered clusters of particles vibrated in liquid were studied in
Ref. \cite{voth}.

\begin{figure}[ptb]
\includegraphics[angle=0,width=3in]{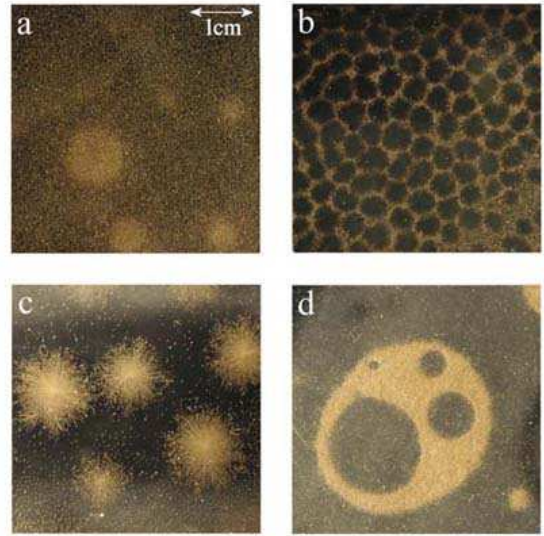}
\caption{Snapshots of a) static clusters; b) honeycomb precipitate
c) "down" toroidal vortices; d) pulsating rings.} \label{figure1}
\end{figure}

\begin{figure}[ptb]
\includegraphics[angle=0,width=2.7in]{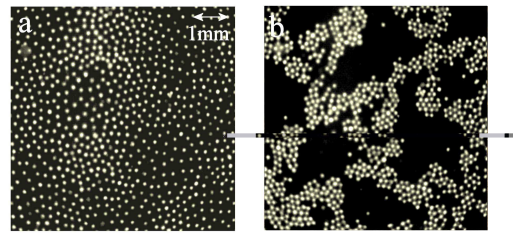}
\caption{Close-up view of (a) Wigner crystal precipitate and (b)
honeycomb precipitate}\label{figure2}
\end{figure}

In this Letter we report new dynamic phenomena occurring in
granular gas in poorly conducting liquid subject to strong
electric field (up to 20 kV/cm). We show that metallic particles
(120 $\mu$m diameter Bronze spheres) immersed in a toluene-ethanol
mixture in DC electric field self-assemble into  a rich variety of
novel phases. These phases include static precipitates: honeycombs
and Wigner crystals; and novel dynamic condensates: toroidal
vortices and pulsating rings (Figs. \ref{figure1} and
\ref{figure2}). The observed phenomena are attributed to
interaction  between particles and EHD flows produced by the
action of the electric field on ionic charges in the bulk of
liquid. This provides a new mechanism for self-assembly of
microparticles in non-aqueous solutions.

To form the electro-cell,  granular media consisting of 3 g (about
$0.5\times10^6$) mono-dispersed 120 $\mu$m bronze spheres was
placed into a 1.5 mm gap between two horizontal 12.5$\times$12.5
cm glass plates covered by a transparent conducting layer of
indium tin-dioxide (the particles constitute less than a monolayer
coverage on the bottom plate).  An electric field perpendicular to
the plates was created by a DC high voltage source (0-3 kV)
connected to the inner surface of each plate \cite{ar1,ar2}. The
liquid was introduced into the cell through two teflon
micro-capillaries connected to opposing side walls of the cell.
Real time images were acquired using a high speed, up to 1000
frames per second, digital camera suspended over the transparent
glass plates of the cell. Experiments were repeated with smaller
particles (40 $\mu$m Cu spheres) and qualitatively similar results
were obtained.

The principle of the electro-cell is as follows:  a particle
acquires an electric charge when it is in contact with the bottom
conducting plate. It then experiences a force from the electric
field between the plates.  If the upward force induced by the
electric field exceeds gravity, the particle travels to the upper
plate, reverses charge upon contact, and is repelled down to the
bottom plate. This process repeats in a cyclical fashion. In an
air-filled cell, the particle remains immobile at the bottom plate
if the electric field $E$ is smaller than a first critical field
$E_1$. If the field is larger than a second critical field value,
$E_2> E_1$, the system of particles transforms into a gas-like
phase. When the field is decreased below $E_2 $($E_1<E<E_2$)
localized clusters of immobile particles spontaneously nucleate to
form a static precipitate on the bottom plate. The clusters
exhibit the Ostwald-type ripening \cite{meerson}. The phase
diagram for the air-filled cell is symmetric with respect to the
reversal of the electric field direction.

\begin{figure}[ptb]
\includegraphics[angle=0,width=3in]{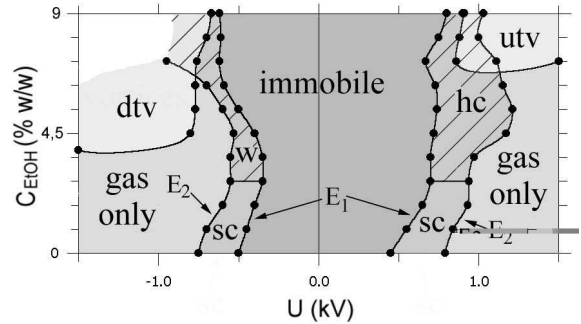}
\caption{Phase diagram, $U$ is applied voltage (positive values
correspond to plus on top plate), $C$ is concentration of ethanol.
Domain sc denotes "static clusters", w-Wigner crystal,
hc-honeycombs, utv and dtv up/down toroidal vortices.
}\label{figure3}
\end{figure}

In this work, we discovered that the situation is remarkably
different when the cell is filled with a non-polar low-viscosity
liquid. For the liquid, we typically used HPLC grade toluene with
various concentrations of ethanol (98\% w/w anhydrous ethanol +
2\% w/w water). The conductivity of the liquid changes from
$5\times10^{-11}$ Ohm$^{-1}$ m$^{-1}$ (pure toluene) to $5\times
10^{-9}$ Ohm$^{-1}$ m$^{-1}$ (9\% ethanol solute). The phase
diagram delineating the different particle behavior as a function
of the applied voltage and concentration of ethanol is shown in
Fig. \ref{figure3}.  For relatively low concentrations of ethanol
solute (below 3\%), the qualitative behavior of the "liquid-filled
cell" is not very different from that of the "air-filled cell":
clustering of immobile particles and coarsening are observed
between two critical field values $E_{1,2}$ with the clusters
being qualitatively similar to that of the "air cell".  A typical
low-concentration pattern is shown in Fig. \ref{figure1}a.

When the ethanol concentration is increased, the behavior of the
liquid-filled cell is remarkably different from that of the
air-filled cell.  The phase boundaries are no longer symmetric
with respect to reversal of the applied voltage. Initially, an
increase in ethanol concentration leads to asymmetric behavior
with respect to the direction of the electric field.  Critical
field values, $E_{1,2}$, are larger when  the electric field is
directed downward ("+" on upper plate) and smaller when the field
is directed upward ("-" on upper plate). This difference grows
with increasing ethanol concentration.

The observed asymmetry of the critical fields is apparently due to
an excess negative charge in the bulk of the liquid. This  charge
increases the effective electric field acting on the
positively-charged particles sitting on the bottom plate when a
positive voltage is applied to that plate and correspondingly
decreases the effective electric field in the case of an applied
negative voltage. From the shift in the critical field values one
estimates the excess negative charge concentration in the bulk.
Assuming uniform charge distribution in the bulk one derives the
estimate for the excess charge concentration $n\approx\varepsilon
\varepsilon_0 \Delta U/ed^2$, where the dielectric permeability of
the toluene/ethanol mixture $\varepsilon \approx 3$, $d$ is the
gap between the plates, and $e$ is ionic charge. For the solute
concentration  $C$=3 \% and the observed critical voltage shift
$\Delta U=150$ V, one finds  $n^- - n^+ \sim 10^{16}$ m$^{-3}$.
The carrier concentration can be independently estimated from
transport measurement (total current through the cell is $I \sim
1$ $\mu$A at $U = 300$ V, its surface is $S=1.5\times 10^{-2}$
m$^2$). Neglecting concentration gradients, the migration current
can evaluated as $e^2DS(n^- + n^+ )U/dk_BT$, where  $D \approx
10^{-9}$ m$^2$/s is the representative value for ion diffusion,
and $T$ is temperature.  That estimate also yields $n^- + n^+ \sim
10^{16}$ m$^{-3}$. Both estimates indicate  that the liquid
contains almost exclusively ions of the same (negative) sign
(albeit in a very small concentration). These   ions (possibly
$OH^-$)  are distributed throughout the cell by the current, while
positive ions ($H^+$) are absorbed by the electrodes
\cite{kitahara}.

The situation changes dramatically for higher ethanol
concentrations: increasing the applied voltage leads to the
formation of two novel immobile phases: honeycomb and Wigner
crystal precipitates. The Wigner crystal precipitate (Fig.
\ref{figure2}a) is induced for upward direction of the applied
electric field, whereas a honeycomb precipitate is induced upon
electric field reversal (Figs. \ref{figure1}b, \ref{figure2}b). A
return to the original applied voltage transforms the honeycomb
structure back into the Wigner crystal phase and vice versa
\cite{foot}. The representative time scale for these transitions
is 0.5 sec. For high enough voltage these patterns coexist with
granular gas. However, due to the bombardment by the "gas- phase"
particles, the honeycomb structure becomes more diffuse.  The
typical size of the elementary honeycomb cell is determined by the
gap between the plates.  Increasing the gap from 1.5 mm to 3 mm
results in doubling of the honeycomb cell size.

\begin{figure}[ptb]
\includegraphics[angle=0,width=2.7in]{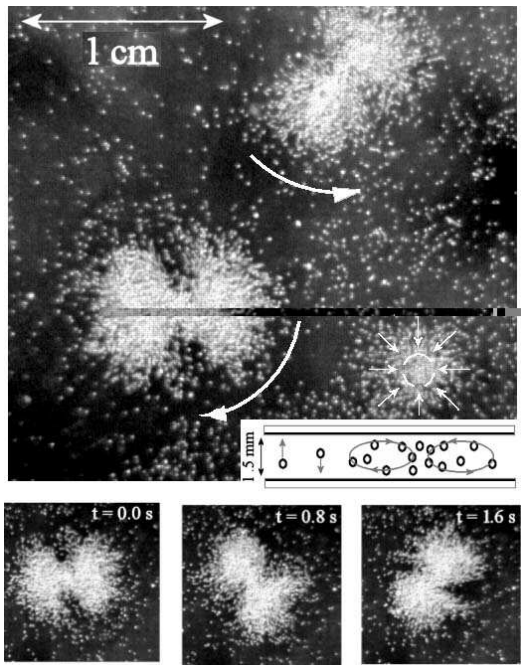}
\caption{Upper panel: two horizontally rotating down t-vortices.
Arrows indicate direction of horizontal rotation of binary
t-vortices. Small arrows show particle trajectories at top plate
for single vortex.  Inset: vertical cross-section of vortex flow
for  down t-vortex. Lower panel: time evolution of one rotating
asymmetric down t-vortex.}\label{figure4}
\end{figure}

A further increase of ethanol concentration leads to the
appearance of a novel dynamic phase - condensate (Fig.
\ref{figure1}c,d). The dynamic condensate is very different from
the static precipitate observed for low ethanol concentrations and
voltages: almost all particles in the condensate are engaged in a
circular vortex motion in the vertical plane, resembling that of
particles in Rayleigh-B\'enard convection, see Fig. \ref{figure4}
\cite{boden}. The condensate co-exists with the dilute granular
gas. The direction of rotation is determined by the polarity of
the applied voltage: particles stream towards the center of the
condensate near the top plate for upward field direction and vice
versa. The direction of rotation is visible using particle-image
velocimetry.  In other words, for upward field direction the
particles stream down at the center of the condensate domain
forming downward toroidal vortices (Fig. \ref{figure1}c).
Conversely, the particles form upward toroidal vortices for
downward field direction.

\begin{figure}[ptb]
\includegraphics[angle=0,width=2.7in]{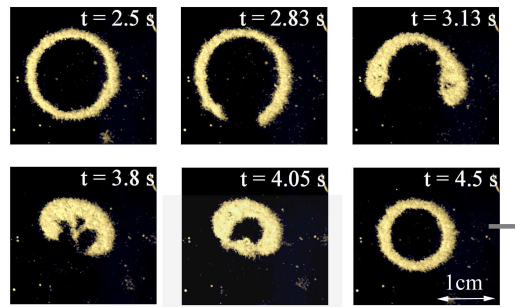}
\caption{Time evolution of a pulsating ring.}\label{figure6}
\end{figure}

The evolution of the condensates depends on the electric field
direction. For downward field, the "up" t-vortices become unstable
due to the spontaneous formation of voids in the condensate (Fig.
\ref{figure1}d). These voids exhibit complex intermittent
dynamics. Medium-size clusters typically form only a single void
and fascinating "pulsating" rings appear (see Fig. \ref{figure6}).
In contrast, for upward field, large "down" t-vortices remain
stable. In some cases two or more "down" t-vortices merge into
one, forming an asymmetric condensate which performs composite
rotation in the horizontal plane, see Fig. \ref{figure4}.

The honeycombs, Wigner crystal, and t-vortices are likely caused
by the same mechanism. It is known that electric field generates
local EHD convective flow in charged liquid in the vicinity of
microparticles \cite{trau,yeh,solom}. We observed that for the
downward field direction the negatively charged liquid flows
upwards around a stationary lying particle, creating an axially
symmetric toroidal flow field. The upward flow induces a sweeping
horizontal flow on the bottom plate toward the center of the
precipitate. Such EHD-induced flow can assemble particles into a
honeycomb structure.  The proportionality of the honeycombs scale
to the gap between plates supports the EHD convective mechanism
for honeycomb formation (compared to Rayleigh-B\'enard
convection). In the case of upward field direction, each particle
induces a downward flow of the fluid and, consequently, a sweeping
horizontal flow away from the center of a precipitate, leading to
mutual repulsion and the formation of a Wigner crystal. For
elevated concentrations of ethanol and higher voltages the EHD
flow is able to drag bouncing particles from the gas phase and
assemble them into t-vortex condensates.  This mechanism is
consistent with the observation that particles stream outwards
from the center of a "down" t-vortex on the bottom plate (as in
the Wigner crystal) and inward towards the center, of an "up" t-
vortex (as in honeycombs). Spontaneously-formed voids in large,
"up" t-vortices are inflated by EHD flow (like voids in a
honeycomb structure), leading to persistent modification of the
shape of the pulsating rings.

\begin{figure}[ptb]
\includegraphics[angle=0,width=2.5in]{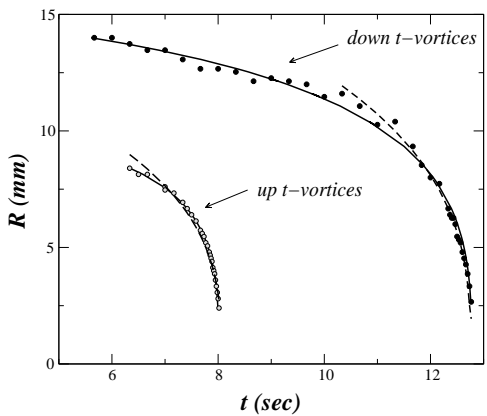}
\caption{Distance $R$ between two vortices vs time $t$. Symbols
show experimental data, solid lines are fits according to Eq.
\protect (\ref{eq1}), dashed lines are $R\sim (t_0-t)^{1/3}$, for
$U=-800$ V, $C=8\%$ w/w  (dtv) and $U=1900$ V, $C=8\%$ w/w
(utv).}\label{figure5}
\end{figure}

The toroidal vortices (t-vortices) attract each other and
coalesce, resulting in large-scale dynamic condensates. The
representative results illustrating hydrodynamic attraction
between t-vortices are shown in Fig. \ref{figure5}. The separation
between two t-vortices, $R$, for small distances with respect to
the cell gap $h$ agrees with the following power law $R \sim
(t_0-t)^{1/3}$, where $t_0=const$. This behavior yields the
following expression for the attraction velocity $V=dR/dt =
-\kappa /R^2$, $\kappa=const$ which  strongly indicates the
hydrodynamic character of the vortex interaction. Indeed, in our
quasi-two-dimensional geometry, the asymptotic hydrodynamic
velocity $v$ for inviscous flow far away from the vortex  behaves
as $v \sim 1/R$. Due to incompressibility $\int v_{\parallel}(z)
dz =0$, where $z$ is the vertical coordinate and $v_{\parallel}$
is horizontal hydrodynamic velocity. Therefore, in the lowest
order, the interaction between  t-vortices is determined by the
square of the hydrodynamic velocity $v_{\parallel}$, i.e. $V \sim
v_{\parallel}^2 \sim -1/R^2$. In the presence of other t-vortices
and due to influence of the boundaries the interaction law is
modified as follows
\begin{equation}
\frac{dR}{dt}= -\frac{\zeta V_0}{R} -\frac{\kappa}{R^2} + O(V_0^2)
\label{eq1}
\end{equation}
where $V_0= \sum_i  v_{\parallel i}$ is mean hydrodynamic velocity
created by other vortices and $ \zeta=const$. Assuming $V_0\approx
const$ (which is true when all other vortices are far away from
the two vortices of the interest) one obtains excellent fitting
for the experimental curve in Fig. \ref{figure5}.

An order-of-magnitude estimate of EHD velocity can be obtained
from the balance of viscous ($\eta \nabla^2v$) and electric
($enE$) forces acting on the charged liquid. Here
$\eta=5.6\times10^{-4}$ Pa$\cdot$s is the viscosity of toluene and
$n$ is the ion concentration. Since the velocity varies on a
characteristic scale of the order of the particle radius $a$, one
can replace $\nabla^2$ with $a^{-2}$, and thus $v \approx
enEa^2/\eta$. Taking $n \sim  10^{16}$ m$^{-3}$ gives an EHD
velocity of $\sim 5$ mm/sec at an applied voltage $U=500$ V. This
estimate is in qualitative agreement with the measured times of
pattern formation and vortex attraction.

We also carried out experiments with solutions of tributylamine
tetraphenilborate (TBATPB) in toluene. This system shows the same
phases but with opposite polarity of the applied voltage. The
critical lines in the phase diagram of the TBATPB/toluene mixture
have slopes opposite to that of the ethanol/toluene based phase
diagram (see Fig. \ref{figure3}). This observation depicting
reversal of behavior under opposite polarity indicates the
presence of positive ions in the bulk of the solution.

There is a significant body of research on the behavior of {\it
dielectric} colloidal particles in aqueous solutions [see e.g.
Refs. \cite{hayward,trau,yeh,solom}]. However, dynamic patterns
were not observed. The main difference here is the presence of
thick spatially-charged layer in the dielectric liquid. The fixed
sign excess charge in the bulk results in a different kind of
interaction between particles and the EHD flows.

In conclusion, we report the discovery of a series of remarkably
rich novel vortex phases and the dynamic assembly of electric
field-driven conductive particles in poorly conductive fluids. Our
results exemplify the importance of interactions between the
particles and uncompensated ionic charges in the bulk of the
liquid. Extension of our experiments toward self-assembly of
submicron particles may have intriguing practical applications,
such as lithographic patterning and assembly
\cite{hayward,yeh,deng}, etc. We are grateful to A.A. Abrikosov
and V.M. Vinokur for fruitful discussions. This research was
supported by the US DOE, grant W-31-109-ENG-38.

\end{document}